\documentclass[floatfix, noshowpacs, preprintnumbers, twocolumn, amsmath, amssymb, aps, superscriptaddress, prb]{revtex4}

\pdfoutput=1

\usepackage{graphicx, xcolor}
\usepackage[caption=false]{subfig}
\usepackage{soul}

\begin{document}

\title{Quantum suppression law in a 3-D photonic chip\\ implementing the Fast Fourier Transform}

\author{Andrea Crespi}
\affiliation{Istituto di Fotonica e Nanotecnologie, Consiglio Nazionale delle Ricerche (IFN-CNR), 
Piazza Leonardo da Vinci, 32, I-20133 Milano, Italy}
\affiliation{Dipartimento di Fisica, Politecnico di Milano, Piazza Leonardo da Vinci, 32, I-20133 Milano, Italy}

\author{Roberto Osellame}
\email{roberto.osellame@polimi.it}
\affiliation{Istituto di Fotonica e Nanotecnologie, Consiglio Nazionale delle Ricerche (IFN-CNR), 
Piazza Leonardo da Vinci, 32, I-20133 Milano, Italy}
\affiliation{Dipartimento di Fisica, Politecnico di Milano, Piazza
Leonardo da Vinci, 32, I-20133 Milano, Italy}

\author{Roberta Ramponi}
\affiliation{Istituto di Fotonica e Nanotecnologie, Consiglio Nazionale delle Ricerche (IFN-CNR), 
Piazza Leonardo da Vinci, 32, I-20133 Milano, Italy}
\affiliation{Dipartimento di Fisica, Politecnico di Milano, Piazza Leonardo da Vinci, 32, I-20133 Milano, Italy}

\author{Marco Bentivegna}
\affiliation{Dipartimento di Fisica, Sapienza Universit\`{a} di Roma,
Piazzale Aldo Moro 5, I-00185 Roma, Italy}

\author{Fulvio Flamini}
\affiliation{Dipartimento di Fisica, Sapienza Universit\`{a} di Roma,
Piazzale Aldo Moro 5, I-00185 Roma, Italy}

\author{Nicol\`o Spagnolo}
\affiliation{Dipartimento di Fisica, Sapienza Universit\`{a} di Roma,
Piazzale Aldo Moro 5, I-00185 Roma, Italy}

\author{Niko Viggianiello}
\affiliation{Dipartimento di Fisica, Sapienza Universit\`{a} di Roma,
Piazzale Aldo Moro 5, I-00185 Roma, Italy}

\author{Luca Innocenti}
\affiliation{Dipartimento di Fisica,  Universit\`{a} di Roma Tor Vergata,
Via della ricerca scientifica 1, I-00133 Roma, Italy}

\author{Paolo Mataloni}
\affiliation{Dipartimento di Fisica, Sapienza Universit\`{a} di Roma,
Piazzale Aldo Moro 5, I-00185 Roma, Italy}

\author{Fabio Sciarrino}
\email{fabio.sciarrino@uniroma1.it}
\affiliation{Dipartimento di Fisica, Sapienza Universit\`{a} di Roma,
Piazzale Aldo Moro 5, I-00185 Roma, Italy}

\maketitle
\textbf{The identification of phenomena able to pinpoint quantum interference is attracting large interest. Indeed, a generalization of the Hong-Ou-Mandel effect valid for any number of photons and optical modes would represent an important leap ahead both from a fundamental perspective and for practical applications, such as certification of photonic quantum devices, whose computational speedup is expected to depend critically on multiparticle interference. Quantum distinctive features have been predicted for many particles injected into multimode interferometers implementing the Fourier transformation in the Fock space. In this work we develop a scalable approach for the implementation of quantum fast Fourier transform using 3-D photonic integrated interferometers, fabricated via femtosecond laser writing technique. We observe the quantum suppression of a large number of output states with 4- and 8-mode optical circuits: the experimental results demonstrate genuine quantum interference between the injected photons, thus offering a powerful tool for diagnostic of photonic platforms.}

\section{Introduction}
The amplitude interference between wavefunctions corresponding to indistinguishable particles lies at the very heart of quantum mechanics. Right after the introduction of laser amplification, the availability of strong coherent pulses allowed to test interference between different light pulses\cite{Magyar63,Pfleegor67}, while generation of pairs of identical photons through parametric fluorescence \cite{Burnham70} led subsequently to the milestone experiment of Hong, Ou and Mandel\cite{Hong87,Cosme08,Oua88,Walborn03,Sagioro04}. Later on, photonic platforms have been demonstrated to be in principle capable to perform universal quantum computing\cite{KLM01}.

Recently, multi-particle interference effects of many photons in large interferometers are attracting a strong interest, as they should be able to show unprecedented evidences of  the superior quantum computational power compared to that of classical devices \cite{OBri07,Walt05,Walm05}. The main example is given by the Boson Sampling\cite{Aaronson10} computational problem, which consists in sampling from the probability distribution given by the permanents of the $n\times n$ submatrices of a given Haar-random unitary. The problem is computationally hard (in $n$) for a classical computer since calculating the permanent of a complex-valued matrix is a $\#$\textit{P-hard} problem, while it can be efficiently solved by $n$ indistinguishable photons evolving through an optical interferometer implementing the unitary transformation in the Fock space, and detecting output states with an array of single-photon detectors. The chance to provide evidences of a post-classical computation with this relatively simple setup has triggered a large experimental effort, leading to small-scale implementations\cite{Broome2013,Spring2013,Till2012,Crespi2012,Spagnolo2013a,Carolan2013a,Bentivegna2015}, as well as theoretical analyses on the effects of experimental imperfections \cite{Rohde12,Leverrier2013} and on possible implementations including alternative schemes \cite{Motes2013,Rohde2013}.

In the context of searching for experimental evidences against the extended Church-Turing thesis, a Boson Sampling experiment poses a problem of certification of the result's correctness in the computationally-hard regime \cite{Gogolin2013}. The very complexity of the Boson Sampling computational problem precludes the use of a brute-force approach, {\it i.e.} calculating the expected probability distribution at the output and comparing it with the collected data. Efficient statistical techniques able to rule out trivial alternative distributions have been proposed \cite{Aaronson13} and tested\cite{Spagnolo2013a,Carolan2013a}, but the need for more stringent tests able to rule out less trivial distributions has led, and continues to encourage, additional research efforts in this direction.

In particular, an efficient test able to confirm true $n$-photon interference in a multimode device has been recently proposed\cite{Tichy2013}. The protocol relies on the suppression of specific output configurations\cite{Tichy2010} in an interferometer implementing an $n^p$-dimensional Quantum Fourier Transform (QFT) matrix, with $p$ being any integer. QFT is the Discrete Fourier Transform (DFT) applied to the vector of amplitudes of a quantum state, and hence its matrix representation is also called Fourier matrix. This suppression effect is due to granular\cite{Tichy2013} many-particle interference and is thus able to rule out alternative models requiring only coarse-grained effects like the ones present in Bose-Einstein condensates\cite{Cennini05,Hadzibabic2010,Tichy2012}. Indeed, the implications of this effect go well beyond the certification of Boson Sampling devices. As a generalization of the 2-photon/2-modes Hong-Ou-Mandel (HOM) effect, the quantum suppression law is important at a fundamental level, while at the practical level it could be used as a diagnostic tool for a wide range of photonic platforms\cite{Tichy2013,Tichy2010,Peruzzo2010,Crespi2013}.

\begin{figure*}[ht!]
\includegraphics[width=0.8\textwidth]{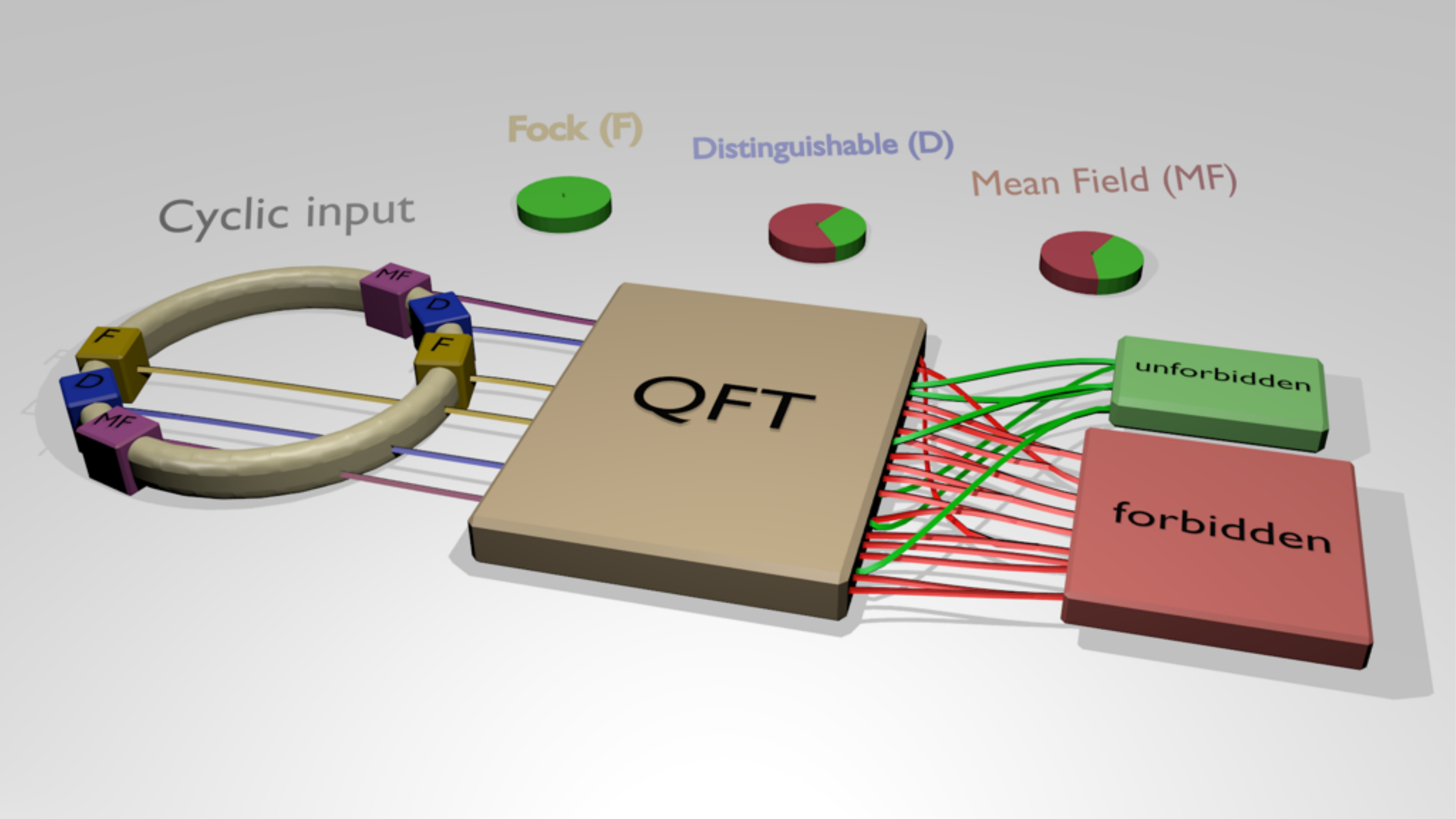}
\caption{\textbf{Quantum suppression law for Fock states in a quantum Fourier transform.} Conceptual scheme of the protocol: the possible configurations of $n$ photons at the output of an $m$-mode interferometer can be divided into two categories, unforbidden and forbidden, depending on whether they satisfy or not the suppression condition (\ref{suppression}), respectively. The pie charts show the expected output statistics with different cyclic input states, {\it i.e.} composed by different classes of particles, where green and red areas represent events with unforbidden and forbidden outputs, respectively. The injection of a cyclic Fock state in an $m$-mode QFT interferometer results in total suppression of forbidden output states. Cyclic states with distinguishable particles show no suppression, so that each output combination is equally likely to occur. A Mean Field state, which reproduces some of the features of bosonic statistics by single particle interference and phase averaging, shows suppression with highly reduced contrast.}
\label{fig:QFTConcept}
\end{figure*}

In this article, we report the experimental observation of the recently theoretically proposed\cite{Tichy2013} suppression law for QFT matrices, and its use to validate quantum many-body interference against alternative non-trivial hypotheses resulting in similar output probability distributions. QFTs have been implemented with an efficient and reliable approach by exploiting the quantum version of the Fast Fourier Transform, an algorithm optimized to compute the DFT of a sequence. Here we implement Quantum Fast Fourier Transforms (QFFTs) on photonic integrated interferometers, using a theoretical proposal developed by Barak and Ben-Aryeh\cite{Barak2007}, by exploiting the 3-D capabilities of the adopted integrated circuits. The optical circuits are realized by femtosecond laser writing\cite{Osellame2003,gattass2008flm}, which makes it possible to fabricate waveguides arranged in three-dimensional structures with arbitrary layouts\cite{Meany12,Spagnolo13,Poulios14}. The observations have been carried out with two-photon Fock states injected into 4-mode and 8-mode QFFTs. The peculiar behaviour of Fock states compared to other kinds of states is investigated, showing in principle the validity of the certification protocol for the identification of real granular $n$-particle interference, which is the source of a rich landscape of quantum effects such as the computational complexity of Boson Sampling.

\begin{figure*}[ht!]
\includegraphics[width=1\textwidth]{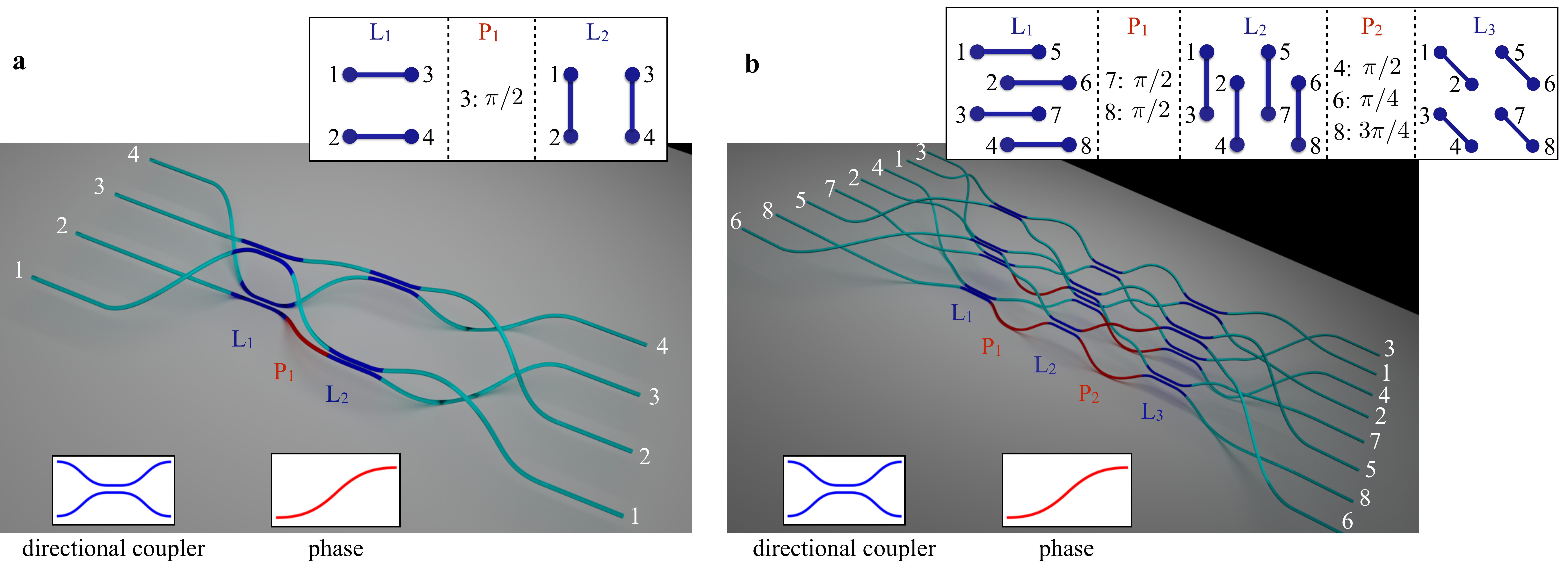}
\caption{\textbf{Schematic of the structure of the integrated devices.} Internal structure of the 4-mode ({\bf a}) and 8-mode ({\bf b}) integrated interferometers implementing the QFFT in the Fock space. In the 8-mode case, the Barak and Ben-Aryeh algorithm requires an additional relabeling of the output modes (not shown in the figure), namely $2 \leftrightarrow 5$ and $4 \leftrightarrow 7$, to obtain the effective Fourier transformation. The mode arrangement has been chosen in a way to minimize bending losses. The insets show the actual disposition of the waveguides in the cross-section of the devices. The modes coupled together in each step ($L_{i}$) of the interferometer are joined by segments. The implemented phase shifts in each step ($P_{i}$) are also indicated.}
\label{fig:strutturaChip}
\end{figure*}

\subsection{Quantum suppression law in Fourier transform matrices}
As a generalization of the HOM effect, it has been pointed out that quantum interference effects in multimode interferometers may determine suppression of a large fraction of the output configurations\cite{Tichy2010, Tichy2012, Crespi2015}, depending on the specific unitary transformation being implemented and on the symmetry of the input state. In particular\cite{Tichy2010, Tichy2012}, let us consider a cyclic input, {\it i.e.} an $n-$photon Fock state over $m=n^p$ modes (for some integer $p$) where the occupied modes $j_{r}^s$ are determined by the rule $j_{r}^s=s+(r-1) \, n^{p-1}$, with  $r=1, \ldots, n$ and $s=1, \ldots, n^{p-1}$. The index $s$ takes into account the fact that there are $n^{p-1}$ possible $n$-photon arrangements with periodicity $n^{p-1}$, which simply differ by a translation of the occupational mode labels. For example, for $n=2$ and $m=4$ there are $2^1=2$ possible cyclic states, (1,0,1,0) and (0,1,0,1), while for $n=2$ and $m=8$ there are $2^2=4$ possible (collision-free) cyclic inputs, {\it i.e.} the states (1,0,0,0,1,0,0,0), (0,1,0,0,0,1,0,0), (0,0,1,0,0,0,1,0) and (0,0,0,1,0,0,0,1).

We consider the evolution of such states through an interferometer implementing the QFT over the Fock space, described by the unitary matrix
\begin{equation}
U_{l,q}^{\operatorname{QFT}}=\frac{1}{\sqrt{m}}\, e^{i\, \frac{2 \pi l q}{m}}.
\label{fourier:formula}
\end{equation}

Such evolution results in the suppression of all output configurations not fulfilling the equation

\begin{equation}
\mod \!\!\left(\sum_{l=1}^n k_l,n\right ) = 0,
\label{suppression}
\end{equation}
 
where $k_l$ is the output mode of the $l^{th}$ photon. An interesting application of suppression laws is to certify the presence of true many-body granular interference during the evolution in the interferometer, ruling out alternative hypotheses which would result in similar output probability distributions. In particular, in the case of QFT matrices, the observation of the suppression law (2) allows to certify that the sampled output distribution is not produced by neither distinguishable particles nor a Mean Field state (MF)\cite{Tichy2013}. The latter is defined as a single particle state $|\psi^s\rangle$ of the form

\begin{equation}
|\psi^s\rangle = \frac{1}{\sqrt{n}}\sum_{r=1}^n e^{i\theta_r}| j_r^s \rangle,
\end{equation}

with a random set of phases $\theta_r$ for each state, being $| j_r^s \rangle$ a single-particle state occupying input mode $j_{r}^s$. This state reproduces macroscopic interference effects, such as bunching or boson clouding\cite{Carolan2013a}, which makes it an optimal test bed for the certification protocol based on the quantum suppression law (see Fig.\ref{fig:QFTConcept}). 

It is possible to quantify the degree of violation $\mathcal D = N_{\operatorname{forbidden}}/N_{\operatorname{events}}$
of the suppression law as the number of observed events in forbidden output states divided by the total number of events\cite{Tichy2013}. If a Fock state was injected in an interferometer implementing the QFT with no experimental imperfections, a violation $\mathcal D= 0$ would be observed. In the case of distinguishable photons there is no suppression law, and the violation will be simply the fraction of suppressed outputs, each one weighted with the number of possible arrangements of the $n$ distinguishable particles in that output combination. In the case of two-photon states, the weighting factor is 2 for collision-free outputs and 1 otherwise, and a degree of violation of $1/2$ is expected (see Sect.~\ref{obs}). On the contrary, in the case of two-photon MF, bunching effects occur leading to an expected degree of violation of half the weighted fraction of suppressed outputs ($1/4$ for two-photon MF). It has been shown that the fraction of forbidden outputs is always large\cite{Tichy2012}. Hence, a comparison of the observed value of $\mathcal D$ with the expected one represents an efficient way, in terms of necessary experimental runs, to discriminate between Fock states, distinguishable particles states and MFs.

\begin{figure*}[ht!]
\includegraphics[width=0.9\textwidth]{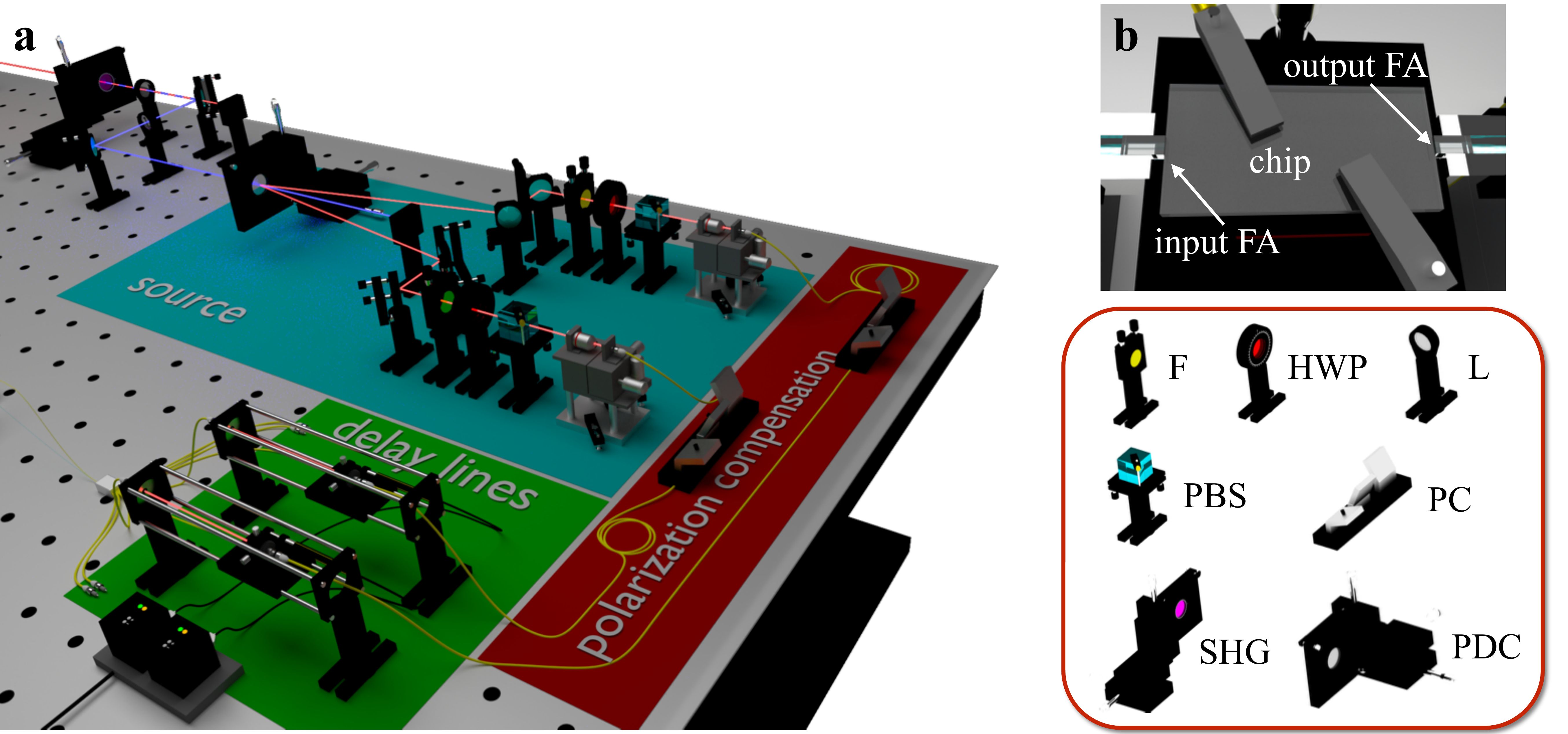}
\caption{\textbf{Experimental apparatus for input state preparation.} {\bf a)} The photon source (F: filter, HWP: half-wave plate, L: lens, PBS: polarizing beamsplitter, PC: polarization compensator, SHG: second harmonic generation, PDC: parametric downconversion). {\bf b)} Photon injection (extraction) before (after) the evolution through the interferometer (FA: fiber array).}
\label{fig:apparato}
\end{figure*}

\subsection{Realization of Quantum Fast Fourier Transform}

Let us now introduce the experimental implementation of the QFFT. The general method to realize an arbitrary unitary transformation  using linear optics was introduced by Reck et al. \cite{Reck1994}, who provided a decomposition of a unitary of dimension $m$ as a sequence of $m(m-1)/2$ beam splitters and phase shifters. However, in the special case of DFT matrices a more efficient method has been proposed \cite{Torma1965,Barak2007}, which takes advantage of the DFT symmetries to significantly reduce the number of linear optical elements required. Based on the classical algorithm of Cooley and Tukey \cite{Cooley1965}, who first introduced the Fast Fourier Transform (FFT) as a more efficient way to calculate the DFT, Barak and Ben-Aryeh developed a quantum analogue in the linear optics domain, leading to the concept of Quantum Fast Fourier Transform (QFFT). This approach, valid for $2^p$-dimensional Fourier matrices, requires only  $(m/2) \log m$ beam splitters and phase shifters, to be compared with the $O(m^2)$ elements needed for the more general Reck decomposition, thus enhancing the compactness and scalability of the platform for a more reliable experimental realization. The overall linear transformation on the optical modes implemented by the QFFT circuit is naturally equivalent to that of the QFT, hence $U^{\operatorname{QFFT}} = U^{\operatorname{QFT}} $.

\section{Results}

\subsection{Realization of 3D QFFT interferometers by femtosecond laser writing}

Here we introduce a new methodology for an integrated implementation of the QFFT which exploits the 3-D capabilities of the femtosecond laser-writing technique.

The sequential structure arising from the decomposition of the $m$-dimensional DFT using the Barak and  Ben-Aryeh algorithm is reproduced by the consecutive layers shown in Fig.\ref{fig:strutturaChip}. The complex arrangement of pairwise interactions necessary for the QFFT algorithm cannot be easily implemented using a planar architecture. However, femtosecond laser writing technique allows to overcome this issue exploiting the third dimension, arranging the waveguides along the bidimensional sections of the integrated chip.

The strategy can be outlined as follows (see also Supplementary Note 1): the $2^p$ modes are ideally placed on the vertices of a $p$-dimensional hypercube; in each step of the algorithm the vertices connected by parallel edges having one specific direction are made to interact by a 2-mode Hadamard transformation, with proper phase terms. An optical interferometer implementing this procedure is thus composed of $\log_2 m = p$ sections, each employing $m/2$ balanced beam splitters and phase shifters. 

We fabricated waveguide interferometers realizing the FFT for $m = 4$ and $m = 8$ modes in borosilicate glass chips using femtosecond laser micromachining\cite{Osellame2003,gattass2008flm}. A schematic representation of these two interferometers is given in Fig. \ref{fig:strutturaChip}. According to the scheme outlined above and by exploiting the three-dimensional capabilites of the fabrication technique, the waveguides are placed, for what concerns the cross section of the device, on the vertices of a 2D projection of the $p$-dimensional hypercube (see also Supplementary S-1). Three-dimensional directional couplers, with proper interaction length and distance to achieve a balanced splitting, connect in each step the required vertices. The insets of Fig. \ref{fig:strutturaChip} show, at each step $i$, which modes are connected by directional couplers  ($L_i$) and the amount of phase shift that needs to be introduced in specific modes ($P_i$). Phase shifters, where needed, are implemented by geometrical deformation of the connecting S-bends. Fan-in and fan-out sections at the input and output of the devices allows interfacing with 127-$\mu$m spaced single-mode fiber arrays. 
Note that in our device geometry, in each step, the vertices to be connected are all at the same relative distance: this means that, unless geometric deformations are designed where needed, light travelling in different modes does not acquire undesired phase delays. Finally, it is worth noting that the geometric construction here developed is scalable
to an arbitrary number of modes.

\subsection{Photon generation and measurement}

The two implemented interferometers of $m=4$ and $m=8$ modes are injected by single-photon and two-photon states. Photons are generated by a type II parametric down-conversion process occuring in a BBO (beta-barium-borate) crystal, pumped by a Ti:Sa pulsed laser (Fig.  \ref{fig:apparato}). Spectral and spatial selection on the two photons are performed by means of interferential filters and single mode fibers respectively. The indistinguishability of the photons is then ensured by a polarization compensation stage, and by propagation through independent delay lines before injection within the interferometer via a single mode fiber array. Delay lines are used to adjust the degree of temporal distinguishability between the photons, so as to record HOM interference patterns. After the evolution occurring through the interferometer, photons are collected by a multimode fiber array. The detection system consists of 4(8) single photon avalanche photodiodes used for the 4- (8-) modes chip. Further details on the photon generation and detection scheme are provided in the Supplementary Information. 

To test the validity of the quantum suppression law, we measured the number of coincidences  at each forbidden output combination injecting cyclic inputs with two indistinguishable photons. We recorded the number of coincidences for each output combination as a function of the temporal delay between the two injected photons, using an electronic data acquisition system able to detect coincidences between all pairs of output modes.

\begin{figure*}[ht!]
\includegraphics[width=0.95\textwidth]{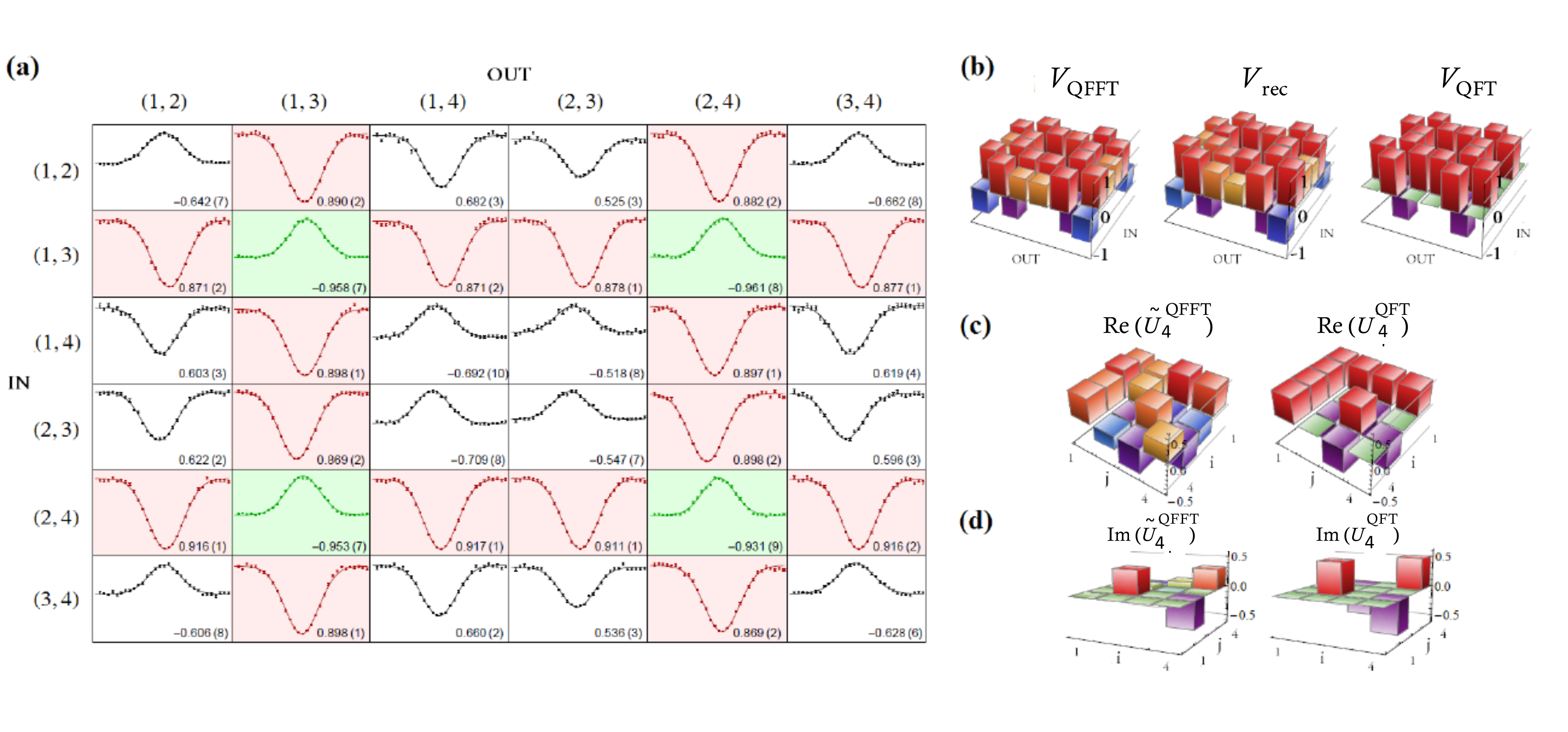}
\caption{\textbf{Quantum suppression law in a 4-mode FFT integrated chip.} {\bf a)} Complete set of 36 visibilities measured for all input-output combinations in the 4-mode chip. For each input-output combination, the measured coincidence pattern as a function of the time delay is shown (points: experimental data, lines: best-fit curves). Cyclic inputs, highlighted in the picture, exhibit enhancement (green) and suppression (red) on cyclic and non-cyclic outputs, respectively. For all points, error bars are due to the poissonian statistics of the events. For each visibility, error bars are obtained through a Monte-Carlo simulation by averaging over 3000 simulated data sets. {\bf b)} HOM visibilities for all 36 input-output configurations. From left to right: experimental measured visibilities ($V_{\operatorname{QFFT}}$), visibilities calculated from the reconstructed unitary ($V_{\operatorname{rec}}$), and visibilities calculated from the theoretical unitary ($V_{\operatorname{QFT}}$). {\bf c-d)} Real and imaginary part of the reconstructed experimental transformation  ($\tilde{U}_4^{\operatorname{QFFT}}$) and of the QFT ($U_4^{\operatorname{QFT}}$).}
\label{fig:Results4}
\end{figure*}

\subsection{One- and two-photon measurements in integrated Fourier interferometers}

Due to the adopted configuration for the type II parametric down-conversion source, which generates two photons in two distinct spatial modes, we limited ourselves to the collision-free input case, in which two different input modes of the chip are injected. The degree of violation $\mathcal D$ of the suppression law could simply be evaluated with a counting experiment. Alternatively, the same quantity $\mathcal D$ can be expressed as a function of single-photon input-output probabilities and of the HOM visibilities, defined as
\[
V_{i,j} = \frac{C_{i,j}-Q_{i,j}}{C_{i,j}}
\]
 where $C_{i,j}$ is the number of detected coincidences for distinguishable photons and $Q_{i,j}$ for indistinguishable photons. The subscripts $(i,j)$ are the indexes of the two output modes, for a given input state. The degree of violation can therefore be expressed as:

\begin{equation}
\begin{aligned}
	\mathcal D &= 
	\frac{N_{\operatorname{forbidden}}}{N_{\operatorname{events}}} =  
	P_{\operatorname{forbidden}} = \\
	&= \sum_{{(i,j)}_{\operatorname{forbidden}}}P_{i,j}^Q = \sum_{{(i,j)}_{\operatorname{forbidden}}}P_{i,j}^C(1-V_{i,j})
\label{violations}
\end{aligned}
\end{equation}

where $P_{i,j}^Q$ ($P_{i,j}^C$) are the probabilities of having photons in the outputs ${i,j}$ in the case of indistinguishable (distinguishable) particles. Here $P_{i,j}^C$ can be obtained from single-particle probabilities.

For the 4-mode device, the full set of $\binom{4}{2}$$^2$= 36 collision-free input-output combinations has been measured by recording the number of coincidences at the two outputs as a function of the temporal delay between the two input photons.  Due to the law given by Eq.(\ref{suppression}), we expect to observe four suppressed outcomes (over six possible output combinations) for the two cyclic input states (1,3) and (2,4). Since distinguishable photons exhibit no interference, HOM dips in the coincidence patterns are expected for the suppressed output states. Conversely, peaks are expected in the non-suppressed output combinations. The experimental results are shown in  Fig.\ref{fig:Results4}-a, where the expected pattern of four suppressions and two enhancements is reported, with average visibilities of $\overline{V}_{\mathrm{dips}} = 0.899 \pm 0.001$ and $\overline{V}_{\mathrm{peaks}} = -0.951 \pm 0.004$ for suppression and enhancement respectively. For the cyclic inputs, we also measured the interference patterns for the output contributions where the two photons exit from the same mode. These cases correspond to a full bunching scenario with $n=2$, and a HOM peak with $V=-1$ visibility is expected independently from the input state and from the unitary operation\cite{Spagnolo13a}. This feature has been observed for the tested inputs, where an average visibility of $\overline{V} = -0.969 \pm 0.024$ has been obtained over all full bunching combinations.

\begin{figure*}[ht!]
\includegraphics[width=0.95\textwidth]{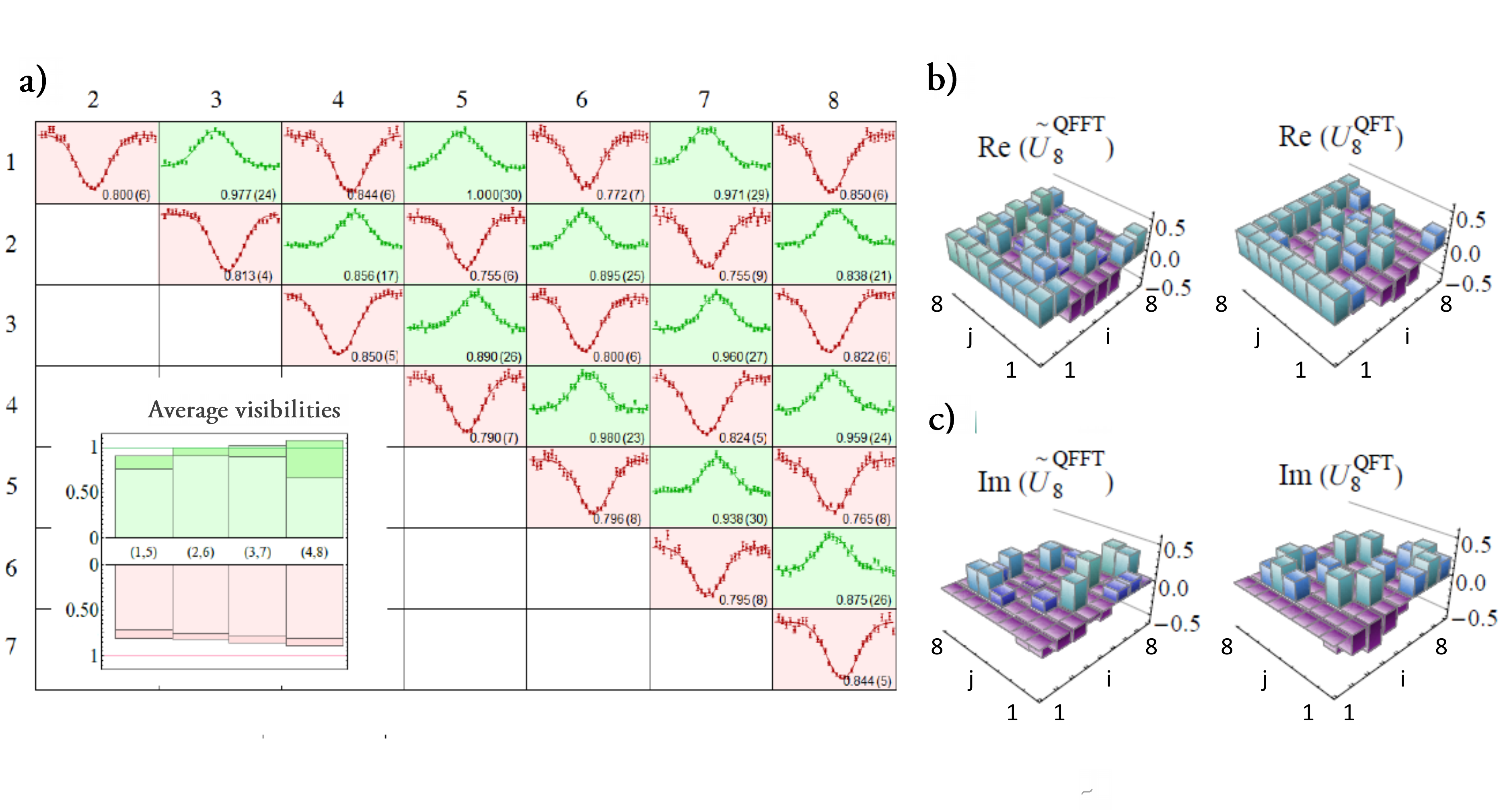}
\caption{\textbf{Quantum suppression law in a 8-mode FFT integrated chip.} {\bf a)} Set of 28 measured visibilities, corresponding to all collision-free output combinations for the input (2,6) of the 8-mode interferometer. For each output combination, the measured coincidence pattern as a function of the time delay is shown (points: experimental data, lines: best-fit curves). Red or green backgrounds correspond to dips and peaks, respectively. For all points, error bars are due to the Poissonian statistics of the events. For each visibility, error bars are obtained through a Monte-Carlo simulation by averaging over 3000 green simulated data sets. Bottom-left inset: average visibilities of dips (red bars) and peaks (green bars) observed for the 4 collision-free cyclic inputs [(1,5), (2,6), (3,7), (4,8)]. Darker regions correspond to error bars of $\pm 1$ standard deviation. {\bf c-d)} Real and imaginary part of the reconstructed experimental transformation  ($\tilde{U}_8^{\operatorname{QFFT}}$) and of the QFT ($U_8^{\operatorname{QFT}}$).}
\label{fig:Results8}
\end{figure*} 

The existence of a general rule for the prediction of suppressed output combinations when injecting a cyclic Fock state in a QFT interferometer is due to the intrinsic symmetry of the problem, as opposed to the general Boson Sampling scenario\cite{Aaronson10}. Suppressed outputs for non-cyclic inputs can be predicted by calculating the permanent of the unitary submatrix given by the intersection of columns and rows of $U^{\operatorname{QFT}}$  corresponding to the occupied input and output modes respectively. The injection of the full set of input states, including the non-cyclic ones,  has been employed for a complete chip reconstruction. The complete set of measured dips and peaks is shown in Fig. \ref{fig:Results4}-a, highlighting the symmetry in the Fourier transform pattern. All single-photon and two-photon measurements described have been used for a reconstruction algorithm\cite{Obrien12} which allows to extrapolate the effective unitary implemented by the interferometer $U_4^{\operatorname{QFFT}}$. Single-photon and two-photon data are exploited to retrieve information on the moduli of the unitary elements and on the unitary phases. The results are shown in Fig. \ref{fig:Results4}-c,d. The fidelity between the reconstructed unitary $\tilde{U}_4^{\operatorname{QFFT}}$ and the theoretical Fourier transform $U_4^{\operatorname{QFT}}$ was $F=$ 0.980 $\pm$ 0.002, thus confirming the high quality of the fabrication process. 

For the 8-mode chip we recorded all the $\binom{8}{2}=28$  two-photon coincidence patterns, as a function of the relative delay between the input photons, for each of the 4 collision-free cyclic inputs and for one non-cyclic input. We used single-photon measurements to extrapolate the effective modulus of the implemented unitary,  while two-photon visibilities allowed us to reconstruct the five fabrication phases (specified in Fig.\ref{fig:strutturaChip}) using a $\chi ^2$ minimization algorithm. The non-cyclic input has been chosen in a way to maximize the sensitivity of the measurements with respect to the five fabrication phases. The results are shown in Fig. \ref{fig:Results8}. The fidelity between the reconstructed unitary $\tilde{U}_8^{\operatorname{QFFT}}$ and the ideal 8-mode Fourier transform $U_8^{\operatorname{QFT}}$ is $F=$ 0.956 $\pm$ 0.005. 

\begin{figure*}
\centering
\includegraphics[width=0.99\textwidth]{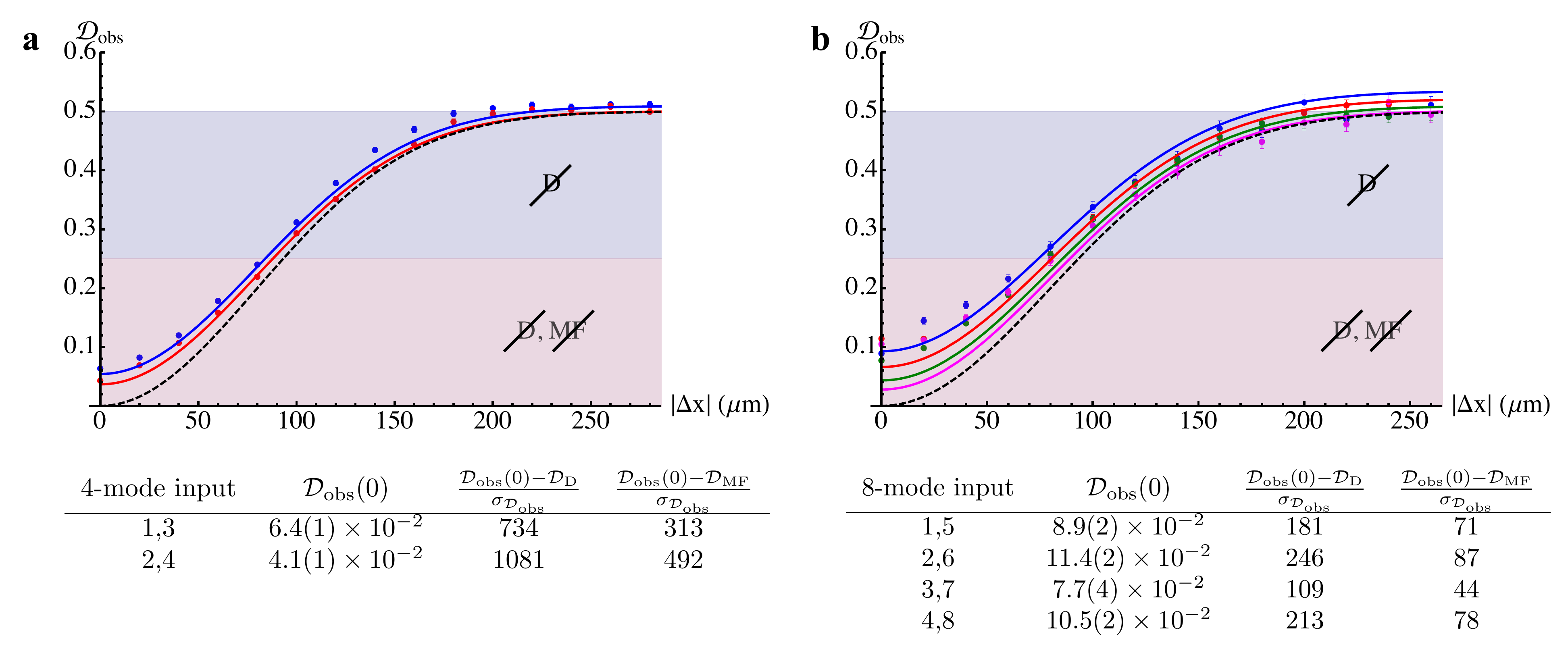}
\caption{{\bf Measured violations.} Observed violations $\mathcal{D}_{\operatorname{obs}}$ as a function of the path difference $\vert \Delta x \vert = c \vert \Delta \tau \vert $ between the two photons. Blue shaded regions in the plots correspond to the cases where the hypothesis of distinguishable particles can be ruled out. Red regions correspond to the cases when both the hypotheses of distinguishable particles and mean field state can be ruled out, and true two-particle interference is present. {\bf a)}, Data for the 4-mode interferometer. Blue points: input (1,3). Red points: input (2,4). Blue solid line: theoretical prediction for input (1,3). Red solid line: theoretical prediction for input (2,4). Black dashed line: theoretical prediction for a QFT matrix. {\bf b)}, Data for the 8-mode interferometer. Blue points: input (1,5). Red points: input (2,6). Green points: input (3,7). Magenta points: input (4,8). Colored solid lines: corresponding theoretical predictions for the different inputs. Black dashed line: theoretical prediction for a QFT matrix. Tables: violations $\mathcal{D}_{\operatorname{obs}}(0)$ at $\Delta x = 0$ and discrepancies (in sigmas) with the expected values for distinguishable particles ($\mathcal{D}_{\operatorname{D}}$) and MFs ($\mathcal{D}_{\operatorname{MF}}$), for the cyclic inputs of the two interferometers. $\mathcal{D}_{\operatorname{obs}}(0)$ are calculated following formula (\ref{violations}), while expected values for the other two cases are $\mathcal{D}_{\operatorname{D}} = 0.5$  and $\mathcal{D}_{\operatorname{MF}}=0.25$. Error bars in all experimental quantities are due to the Poissonian statistics of measured events. All theoretical predictions in solid lines are calculated from the reconstructed unitary and a two-photon state with indistinguishability $\alpha=0.95$.}
\label{fig:violations}
\end{figure*}

\subsection{Observation of the quantum suppression law}
\label{obs}
The suppression of events which do not satisfy Eq. (\ref{suppression}) is fulfilled only when two perfectly indistinguishable photons are injected in a cyclic input of a perfect Fourier transform interferometer. In such a case, we would have the suppression of all output states whose sum of the indexes corresponding to the occupied modes is odd. For the 4-mode and the 8-mode interferometer, this corresponds to 4 (16) suppressed and 2 (12) non-suppressed collision-free outputs (each one given by two possible arrangements of the two distinguishable photons), plus 4 (8) terms with two photons in the same output, each one corresponding to a single possible two-photon path. 

The expected violation for distinguishable particles can be obtained from classical considerations. Let us consider the case with $n=2$. The two distinguishable photons evolve independently from each other, and the output distribution is obtained by classically mixing single-particle probabilities. All collision-free terms are equally likely to occur with probability $q=2/m^2$, while full-bunching events occur with probability $q'= q/2=1/m^2$. The degree of violation $\mathcal D_{\operatorname{D}}$ can then be obtained by multiplying the probability $q$ by the number of forbidden output combinations.
As a result, we expect a violation degree of $\mathcal D_{\operatorname{D}} = 0.5$ for distinguishable two-photon states. The evaluation of the expected value for a mean field state, which is due to single particle bosonic statistic effects, requires different calculations\cite{Tichy2013}. It can be shown that for $n=2$ the degree of violation is $\mathcal D_{\operatorname{MF}} = 0.25$.

For each of the cyclic input, we have evaluated here the violation degree $\mathcal{D}_{\operatorname{obs}}$ resulting from collected data. By measuring the coincidence pattern as a function of the path difference $\Delta x = c \Delta \tau$ between the two photons, and thus by tuning their degree of distinguishability, we could address the transition from distinguishable to indistinguishable particles.
The value of $\mathcal{D}_{\mathrm{obs}}$ as a function of $\Delta x$ has been obtained as $\sum_{(i,j)_\mathrm{forbidden}}P_{i,j}^{\mathrm{C}} (N_{i,j}^{\Delta x}/N_{i,j}^{\mathrm{D}})$, where $N_{i,j}^{\Delta x}$ and $N_{i,j}^{\mathrm{D}}$ are the number of measured coincidences for a given value of $\Delta x$ and for distinguishable particles respectively. Two different regions can be identified. For intermediate values of $\Delta x$ with respect to the coherence length of the photons, the measured data fall below the threshold $\mathcal{D}_{\operatorname{D}}$, and hence the hypothesis of distinguishable particles can be ruled out.
Then, for smaller values of the path difference up to $\Delta x \rightarrow 0$, true two-photon interference can be certified since both hypothesis of distinguishable particles and mean field state can be ruled out. The maximum violation occurring at $\Delta x = 0$ delay can be evaluated using equation (\ref{violations}).
The experimental results retrieved from the protocol are shown in the tables of Fig. \ref{fig:violations},
in which we compare the values $\mathcal{D}_{\operatorname{obs}}(0)$ with the expected values for distinguishable particles $\mathcal{D}_{\operatorname{D}}$ and for a mean field state $\mathcal{D}_{\operatorname{MF}}$.
As shown for our implementation, the robustness of the protocol is ensured by the high number of standard deviation separating the values in each comparison, thus unambiguously confirming the success of the certification protocol.
In conclusion, the alternative hypotheses of distinguishable particles and of a mean field state can be ruled out for all experiments. 

\section{Discussion}

We have reported on the experimental observation of suppression law on specific output combinations of a multimode Fourier transformation in the Fock space due to quantum interference between photons. The observation of the suppression effect allowed to rule out alternative hypotheses to the Fock state. The use of a novel implementation architecture, enabled by the 3-D capabilities of femtosecond laser micromachining, extends the scalability of this technique to larger systems with lower experimental effort with respect to other techniques. At the same time, the universality of a generalized HOM effect with an arbitrary number of particles and modes is expected to make it a pivotal tool in the diagnostic and certification of quantum photonic platforms. Boson Sampling represents a key example, since the scalability of the technique is expected to allow efficient certification of devices outperforming their classical counterparts. 

Fourier matrices can find application in different contexts. For instance, multiport beam splitters described by the Fourier matrix can be employed as building blocks for multiarm interferometers, which can be adopted for quantum-enhanced single and multiphase estimation protocols\cite{Spag12}. This would also allow the measurement of phase gradients with precision lower than the shot-noise limit\cite{Motes15}. Other fields where Fourier matrices are relevant include quantum communication scenarios\cite{Guha11}, observation of two-photon correlations as a function of geometric phase\cite{Laing12}, fundamental quantum information theory including mutually unbiased bases \cite{Beng07}, as well as entanglement generation \cite{Lim05}.

\section*{Methods}

\textbf{Waveguide device fabrication.} Waveguide interferometers are fabricated in EAGLE2000 (Corning Inc.) glass chips. To inscribe the waveguides, laser pulses with 300~fs duration, 220 nJ energy and 1~MHz repetition rate from an Yb:KYW cavity dumped oscillator (wavelength 1030~nm) are focused in the bulk of the glass, using a 0.6 NA microscope objective. The average depth of the waveguides, in the three-dimensional interferometric structure, is 170 $\mu$m under the sample surface. The fabricated waveguides yield single mode behaviour at 800 nm wavelength, with about 0.5 dB~cm$^{-1}$ propagation losses. The central part of the three-dimensional interferometer, which includes all the relevant couplers, have a cross section of about 50~$\mu$m$\times$50~$\mu$m (95~$\mu$m$\times$95~$\mu$m) for a length of 9.0~mm (14.7~mm) in the four-(eight-)modes case. The length of each fan-in and fan-out section, needed to bring the waveguides at 127~$\mu$m relative distance, is 7.8~mm (13.2~mm).

\bigskip

\textbf{Acknowledgements.} 
We acknowledge technical support from Sandro Giacomini and Giorgio Milani. This work was supported by the ERC-Starting Grant 3D-QUEST (3D-Quantum Integrated Optical Simulation; grant agreement no. 307783): http://www.3dquest.eu, by the PRIN project Advanced Quantum Simulation and Metrology (AQUASIM) and by the H2020-FETPROACT-2014 Grant QUCHIP (Quantum Simulation on a Photonic Chip; grant agreement no. 641039).


\begin{thebibliography}{10}
\renewcommand{\bibnumfmt}[1]{#1.}

\bibitem{Magyar63}
\bibinfo{author}{Magyar, G. and Mandel, L.}
\newblock \bibinfo{title}{Interference Fringes Produced by Superposition of Two Independent Maser Light Beams}.
\newblock \emph{\bibinfo{journal}{Nature}}
  \textbf{\bibinfo{volume}{198}}, \bibinfo{pages}{255--256} (\bibinfo{year}{1963}).

\bibitem{Pfleegor67}
\bibinfo{author}{Pfleegor, R. L. and Mandel, L.}
\newblock \bibinfo{title}{Interference of Independent Photon Beams.}
\newblock \emph{\bibinfo{journal}{Phys. Rev. Lett.}}
  \textbf{\bibinfo{volume}{159}}, \bibinfo{pages}{1084--1088} (\bibinfo{year}{1967}).

\bibitem{Burnham70}
\bibinfo{author}{Burnham, D. C. and Weinberg D. L.}
\newblock \bibinfo{title}{Observation of simultaneity in parametric production
of optical photon pairs.}
\newblock \emph{\bibinfo{journal}{Phys. Rev. Lett.}}
  \textbf{\bibinfo{volume}{25}}, \bibinfo{pages}{84--87} (\bibinfo{year}{1970}).

\bibitem{Hong87}
\bibinfo{author}{Hong, C. K., Ou Z. Y.  and Mandel L.}
\newblock \bibinfo{title}{Measurement of subpicosecond time intervals between two photons by interference.}
\newblock \emph{\bibinfo{journal}{Phys. Rev. Lett.}}
  \textbf{\bibinfo{volume}{59}}, \bibinfo{pages}{2046--2016} (\bibinfo{year}{1987}).

\bibitem{Cosme08}
\bibinfo{author}{Cosme, O., Padua, S., Bovino F., Sciarrino, F. and De Martini F.}
\newblock \bibinfo{title}{Hong-Ou-Mandel interferometer with one and two photon pairs.}
\newblock \emph{\bibinfo{journal}{Phys. Rev. A}}
  \textbf{\bibinfo{volume}{77}}, \bibinfo{pages}{053822-1--053822-10} (\bibinfo{year}{2008}).

\bibitem{Oua88}
\bibinfo{author}{Ou, Z. Y. and Mandel, L.}
\newblock \bibinfo{title}{Violation of Bell's Inequality and Classical Probability in a Two-Photon Correlation Experiment.}
\newblock \emph{\bibinfo{journal}{Phys. Rev. Lett.}}
  \textbf{\bibinfo{volume}{61}},  (\bibinfo{year}{1988}).  
  
\bibitem{Walborn03}
\bibinfo{author}{Walborn,S.P., Oliveira, A.N.,Padua,S., and  Monken, C.H.}
\newblock \bibinfo{title}{Multimode Hong-Ou-Mandel interference.}
\newblock \emph{\bibinfo{journal}{Phys. Rev. Lett.}}
  \textbf{\bibinfo{volume}{90}}, \bibinfo{pages}{143601} (\bibinfo{year}{2003}).
  

\bibitem{Sagioro04}
\bibinfo{author}{Sagioro, M.A., Olindo, C., Monken, C.H. and Padua, S.}
\newblock \bibinfo{title}{Time control of two-photon interference.}
\newblock \emph{\bibinfo{journal}{Phys. Rev. A}}
  \textbf{\bibinfo{volume}{69}}, \bibinfo{pages}{053817} (\bibinfo{year}{2004}).


\bibitem{KLM01}
\bibinfo{author}{Knill, E., Laflamme, R., and  Milburn G.J.}
\newblock A scheme for efficient quantum computation with linear optics.
\newblock {\em Nature},  \textbf{\bibinfo{volume}{409}} :46--52 (2001).


\bibitem{OBri07}
\bibinfo{author}{O'Brien, J.L.}
\newblock Optical Quantum Computing.
\newblock {\em Science}, \textbf{318}, 1567-1570 (2007).

\bibitem{Walt05} 
\bibinfo{author}{Walther, P., Resch, K.J., Rudolph, T., Schenck, E., Weinfurter, H., Vedral, V. et al.}
\newblock Experimental one-way quantum computing.
\newblock {\em Nature}, \textbf{434}, 169-176 (2005).

\bibitem{Walm05}
\bibinfo{author}{Walmsley, I.A., Raymer, M.G.}
\newblock Toward Quantum-Information Processing with Photons.
\newblock {\em Science} \textbf{307}, 1733-1734 (2005).


\bibitem{Aaronson10}
\bibinfo{author}{Aaronson, S.} \& \bibinfo{author}{Arkhipov, A.}
\newblock \bibinfo{title}{The computation complexity of linear optics}.
\newblock In \emph{\bibinfo{booktitle}{Proceedings of the 43rd annual ACM
  symposium on Theory of computing}}, \bibinfo{pages}{333--342} (\bibinfo{year}{ACM Press, 2011}).

\bibitem{Broome2013}
\bibinfo{author}{Broome, M.~A.} \emph{et~al.}
\newblock \bibinfo{title}{Photonic Boson Sampling in a tunable circuit}.
\newblock \emph{\bibinfo{journal}{Science}} \textbf{\bibinfo{volume}{339}},
  \bibinfo{pages}{794--798} (\bibinfo{year}{2013}).

\bibitem{Spring2013}
\bibinfo{author}{Spring, J.~B.} \emph{et~al.}
\newblock \bibinfo{title}{Boson Sampling on a photonic chip}.
\newblock \emph{\bibinfo{journal}{Science}} \textbf{\bibinfo{volume}{339}},
  \bibinfo{pages}{798--801} (\bibinfo{year}{2013}).

\bibitem{Till2012}
\bibinfo{author}{Tillmann, M.} \emph{et~al.}
\newblock \bibinfo{title}{Experimental Boson Sampling}.
\newblock \emph{\bibinfo{journal}{Nature Photon.}}
  \textbf{\bibinfo{volume}{7}}, \bibinfo{pages}{540--544} (\bibinfo{year}{2013}).

\bibitem{Crespi2012}
\bibinfo{author}{Crespi, A.} \emph{et~al.}
\newblock \bibinfo{title}{Integrated multimode interferometers with arbitrary designs for photonic boson sampling. }.
\newblock \emph{\bibinfo{journal}{Nature Photon.}}
  \textbf{\bibinfo{volume}{7}}, \bibinfo{pages}{545--549} (\bibinfo{year}{2013}).

\bibitem{Spagnolo2013a}
\bibinfo{author}{Spagnolo N., Vitelli C., Bentivegna M., Brod, D.J.,  Crespi, A.,  Flamini, F.,
  Giacomini S., Milani, G., Ramponi, R.,  Mataloni, P., Osellame, R.,  Galvao, E.F.
  and Sciarrino, F.}
\newblock Experimental validation of photonic boson sampling.
\newblock {\em Nature Photonics},  \textbf{\bibinfo{volume}{8}} :615--620, (2014).

\bibitem{Carolan2013a}
\bibinfo{author}{Carolan, J.,  Meinecke, J.D.A.,  Shadbolt, P.J.,  Russell, N.J., Ismail, N., Worhoff, K., Rudolph, T.,  Thompson, M.G.,  O'Brien, J.L.,  Matthews, J.C.F., and
  Laing, A.}
\newblock On the experimental verification of quantum complexity in linear
  optics.
\newblock {\em Nature Photonics},  \textbf{\bibinfo{volume}{8}} :621--626, (2014).

\bibitem{Bentivegna2015}
\bibinfo{author}{M. Bentivegna, N. Spagnolo, C. Vitelli, F. Flamini, N. Viggianiello, L. Latmiral,
P. Mataloni, D. J. Brod, E. F. Galvão, A. Crespi, R. Ramponi, R. Osellame and F. Sciarrino.}
\newblock Experimental scattershot boson sampling.
\newblock {\em Sci. Adv.},  \textbf{\bibinfo{volume}{1}}, e1400255 (2015).


\bibitem{Rohde12}
\bibinfo{author}{Rohde, P.~P.} \& \bibinfo{author}{Ralph, T.~C.}
\newblock \bibinfo{title}{Error tolerance of the boson-sampling model for
  linear optics quantum computing}.
\newblock \emph{\bibinfo{journal}{Phys. Rev. A}}, \textbf{\bibinfo{volume}{85}},
  \bibinfo{pages}{022332} (\bibinfo{year}{2012}).

\bibitem{Leverrier2013}
\bibinfo{author}{Leverrier, A.} \& \bibinfo{author}{Garcia-Patron, R.}
\newblock \bibinfo{title}{Analysis of circuit imperfections in BosonSampling}
\newblock \emph{\bibinfo{journal}{ Quantum Inf. Comput.}}, \textbf{\bibinfo{volume}{15}},
  \bibinfo{pages}{0489-0512} (\bibinfo{year}{2015}).

\bibitem{Motes2013}
\bibinfo{author}{Motes, K.~R.}, \bibinfo{author}{Dowling, J.~P.} \&
  \bibinfo{author}{Rohde, P.~P.}
\newblock \bibinfo{title}{Spontaneous parametric down-conversion photon sources
  are scalable in the asymptotic limit for boson-sampling}.
\newblock \emph{\bibinfo{journal}{Phys. Rev. A}}, \textbf{\bibinfo{volume}{88}},
  \bibinfo{pages}{063822} (\bibinfo{year}{2013}).

\bibitem{Rohde2013}
\bibinfo{author}{Rohde, P.~P.}, \bibinfo{author}{Moten, K.~R.} \&
  \bibinfo{author}{Dowling, J.~P.}
\newblock \bibinfo{title}{Sampling generalized cat states with linear optics is
  probably hard}.
\newblock \bibinfo{note}{Preprint at http://lanl.arxiv.org/abs/1310.0297 (2013)}.

\bibitem{Gogolin2013}
\bibinfo{author}{Gogolin, C.}, \bibinfo{author}{Kliesch, M.},
  \bibinfo{author}{Aolita, L.} \& \bibinfo{author}{Eisert, J.}
\newblock \bibinfo{title}{Boson-sampling in the light of sample complexity}.
\newblock \bibinfo{note}{Preprint at http://lanl.arxiv.org/abs/1306.3995 (2013).}

\bibitem{Aaronson13}
\bibinfo{author}{Aaronson, S.} \& \bibinfo{author}{Arkhipov, A.}
\newblock \bibinfo{title}{Bosonsampling is far from uniform}.
\newblock \bibinfo{note}{Preprint at http://lanl.arxiv.org/abs/1309.7460 (2013)}.

\bibitem{Tichy2013}
\bibinfo{author}{Tichy, M.~C.}, \bibinfo{author}{Mayer, K.},
  \bibinfo{author}{Buchleitner, A.} \& \bibinfo{author}{Molmer, K.}
\newblock \bibinfo{title}{Stringent and efficient assessment of Boson-Sampling devices}.
\newblock \emph{\bibinfo{journal}{Phys. Rev. Lett.}}
  \textbf{\bibinfo{volume}{113}}, \bibinfo{pages}{020502}
  (\bibinfo{year}{2014}).

\bibitem{Tichy2010}
\bibinfo{author}{Tichy, M.~C.}, \bibinfo{author}{Tiersch, M.},
  \bibinfo{author}{De Melo, F.}, \bibinfo{author}{Mintert, F.} \& \bibinfo{author}{Buchleitner, Andreas}
\newblock \bibinfo{title}{Zero-Transmission Law for Multiport Beam Splitters}.
\newblock \emph{\bibinfo{journal}{Phys. Rev. Lett.}}
  \textbf{\bibinfo{volume}{104}}, \bibinfo{pages}{220405}
  (\bibinfo{year}{2010}).


\bibitem{Tichy2012}
\bibinfo{author}{Tichy, M.~C.}, \bibinfo{author}{Tiersch, M.},
 \bibinfo{author}{Mintert, F.}  \&  \bibinfo{author}{Buchleitner, A.}
\newblock \bibinfo{title}{Many-particle interference beyond many-boson and many-fermion statistics}.
\newblock \emph{\bibinfo{journal}{New J. Phys.}}
  \textbf{\bibinfo{volume}{14}}, \bibinfo{pages}{093015}
  (\bibinfo{year}{2012}).

\bibitem{Cennini05}
 \bibinfo{author}{Cennini, G.}, \bibinfo{author}{Geckeler, C.},  \bibinfo{author}{Ritt, G.} \&  \bibinfo{author}{Weitz, M.}
\newblock \bibinfo{title}{Interference of a variable number of coherent atomic sources},
\newblock \emph{\bibinfo{journal}{Phys. Rev. A}}
  \textbf{\bibinfo{volume}{72}}, \bibinfo{pages}{051601}
  (\bibinfo{year}{2005}).

\bibitem{Hadzibabic2010}
\bibinfo{author}{Hadzibabic, Z.}, \bibinfo{author}{Stock, S.},
  \bibinfo{author}{Battelier, B.}, \bibinfo{author}{Bretin, V.} \& \bibinfo{author}{Bretin, V.}
\newblock \bibinfo{title}{Interference of an Array of Independent Bose-Einstein Condensates}.
\newblock \emph{\bibinfo{journal}{Phys. Rev. Lett.}}
  \textbf{\bibinfo{volume}{93}}, \bibinfo{pages}{180403}
  (\bibinfo{year}{2004}).

\bibitem{Peruzzo2010}
\bibinfo{author}{Peruzzo A, Lobino M, Matthews J C F, Matsuda N, Politi A, Poulios K, Zhou X-Q, Lahini Y, Ismaili N, Worhoff K, Bromberg Y, Silberberg Y, Thompson M G, O'Brien J L}
\newblock \bibinfo{title}{Quantum walks of correlated photons}
\newblock \emph{\bibinfo{journal}{Science}}
\textbf{\bibinfo{volume}{329}}, \bibinfo{pages}{1500-1503}
(\bibinfo{year}{2010}).

\bibitem{Crespi2013}
\bibinfo{author}{Crespi A, Osellame R, Ramponi R, Giovannetti V, Fazio R, Sansoni L, De Nicola F, Sciarrino F, Mataloni P}
\newblock \bibinfo{title}{Anderson localization of entangled photons in an integrated quantum walk}
\newblock \emph{\bibinfo{journal}{Nature Photonics}}
\textbf{\bibinfo{volume}{7}}, \bibinfo{pages}{322-328}
(\bibinfo{year}{2013}).

\bibitem{Barak2007}
\bibinfo{author} {Barak, R.} \& \bibinfo{author} {Ben-Aryeh, Y.}
\bibinfo{title} {Quantum fast Fourier transform and quantum computation by linear optics}
\newblock \emph{\bibinfo{journal}{J.  Opt. Soc.  Am. B}}
\textbf{\bibinfo{volume}{24}}, \bibinfo{pages}{231--240}
  (\bibinfo{year}{2007}).

\bibitem{Osellame2003}
\bibinfo{author}{Osellame, R.} \emph{et~al.}
\newblock \bibinfo{title}{Femtosecond writing of active optical waveguides with
  astigmatically shaped beams}.
\newblock \emph{\bibinfo{journal}{J.  Opt. Soc.  Am. B}}
  \textbf{\bibinfo{volume}{20}}, \bibinfo{pages}{1559--1567}
  (\bibinfo{year}{2003}).

\bibitem{gattass2008flm}
\bibinfo{author}{Gattass, R.} \& \bibinfo{author}{Mazur, E.}
\newblock \bibinfo{title}{Femtosecond laser micromachining in transparent
  materials}.
\newblock \emph{\bibinfo{journal}{Nature Photon.}}
  \textbf{\bibinfo{volume}{2}}, \bibinfo{pages}{219--225}
  (\bibinfo{year}{2008}).
  
\bibitem{Meany12} 
\bibinfo{author}{Meany T, Delanty M, Gross S, Marshall GD, Steel MJ, Withford MJ.}
\newblock \bibinfo{title}{Non-classical interference in integrated 3D multiports.}
\newblock {\em Opt. Express} \textbf{20}: 26895-26905 (2012).

\bibitem{Spagnolo13} 
\bibinfo{author}{Spagnolo N, Vitelli C, Aparo L, Mataloni P, Sciarrino F, Crespi A et al.}
\newblock \bibinfo{title}{Three-photon bosonic coalescence in an integrated tritter.}
\newblock {\em Nature commun.} {\bf 4}: 1606 (2013).

\bibitem{Poulios14} 
\bibinfo{author}{Poulios K, Keil R, Fry D, Meinecke JD, Matthews JC, Politi A.}
\newblock \bibinfo{title}{Quantum walks of correlated photon pairs in two-dimensional waveguide arrays.}
\newblock {\em Phys. Rev. Lett.} {\bf 112}: 143604 (2014).

\bibitem{Crespi2015} 
\bibinfo{author}{Crespi, A.}
\newblock \bibinfo{title}{Suppression laws for multiparticle interference in Sylvester interferometers.}
\newblock {\em Phys. Rev. A} {\bf 91}: 013811 (2015).


\bibitem{Reck1994} 
\bibinfo{author}{Reck, M.}
\newblock \bibinfo{title}{Experimental realization of any discrete unitary operator.}
\newblock {\em Phys. Rev.Lett.} {\bf 73 (1)}: 5861 (1994).

\bibitem{Torma1965}
\bibinfo{author} {T$\ddot{o}$rm$\ddot{a}$, P.} 
\bibinfo{title} {Beam splitter realizations of totally symmetric mode couplers}
\newblock \emph{\bibinfo{journal}{J. Mod.Opt.}}
\textbf{\bibinfo{volume}{43}}, \bibinfo{pages}{245–251}
  (\bibinfo{year}{1996}).

\bibitem{Cooley1965}
\bibinfo{author} {Cooley, J. W.} \& \bibinfo{author} {Tukey, W.}
\bibinfo{title} {An algorithm for the machine calculation of complex Fourier series}
\newblock \emph{\bibinfo{journal}{Math. Comput.}}
\textbf{\bibinfo{volume}{19}}, \bibinfo{pages}{297-301}
  (\bibinfo{year}{1965}).

\bibitem{Spagnolo13a}
\bibinfo{author}{Spagnolo, N., Vitelli, C., Sansoni, L., Maiorino, E., Mataloni, P., Sciarrino, F., Brod, D. J., Galvao, E. F., Crespi, A., Ramponi, R. and Osellame, R.}
\newblock \bibinfo{title}{General Rules for Bosonic Bunching in Multimode Interferometers}
\newblock {\em Phys. Rev. Lett.} {\bf 113}: 130503 (2013).

\bibitem{Obrien12}
\bibinfo{author}{Laing, A.} \& \bibinfo{author}{O'Brien, J.~L.}
\newblock \bibinfo{title}{Super-stable tomography of any linear optical
  device}.
\newblock \bibinfo{howpublished}{Preprint arXiv:1208.2868v1 [quant-ph]}
  (\bibinfo{year}{2012}).
  
  
\bibitem{Spag12}
\bibinfo{author}{Spagnolo, N., Aparo, L., Vitelli, C., Crespi, A., Ramponi, R., Osellame, R., Mataloni, P. and Sciarrino, F.}
\newblock \bibinfo{title}{Quantum interferometry with three-dimensional
  geometry}.
\newblock \emph{\bibinfo{journal}{Scientific Reports}}
  \textbf{\bibinfo{volume}{2}}, \bibinfo{pages}{862} (\bibinfo{year}{2012}).

  
\bibitem{Motes15}
\bibinfo{author}{Motes, K. R., Olson J. P., Rabeaux E. J., Dowling J. P., Olson S. J. and Rohde P. P.}
\newblock \bibinfo{title}{Linear Optical Quantum Metrology with Single Photons: Exploiting Spontaneously Generated Entanglement to Beat the Shot-Noise Limit}.
\newblock {\em Phys. Rev. Lett.} {\bf 114}: 170802 (2015).

\bibitem{Guha11}
\bibinfo{author}{Guha, S.}
\newblock \bibinfo{title}{Structured Optical Receivers to Attain Superadditive Capacity and the Holevo Limit}.
\newblock {\em Phys. Rev. Lett.} {\bf 106}: 240502 (2011).

\bibitem{Laing12}
\bibinfo{author}{Laing, A., Lawson, T., Mart\'{i}n L\'opez, E. and O'Brien, J. L.}
\newblock \bibinfo{title}{Observation of Quantum Interference as a Function of Berry's Phase in a Complex Hadamard Optical Network}.
\newblock {\em Phys. Rev. Lett.} {\bf 108}: 260505 (2012).


\bibitem{Beng07}
\bibinfo{author}{Bengtsson, I., Brudza, W., Ericsson, A., Larsson, J.-A., Tadej, W. and Zyczkowsky, K.}
\newblock \bibinfo{title}{Mutually unbiased bases and Hadamard matrices
of order six}.
\newblock {\em J. Math. Phys.} {\bf 48}: 052106 (2007).

\bibitem{Lim05}
\bibinfo{author}{Lim, Y. L., and Beige, A.}
\newblock \bibinfo{title}{Multiphoton entanglement through a Bell-multiport beam splitter}
\newblock {\em Phys. Rev. A} {\bf 71}: 062311 (2005).

\end{thebibliography}
\end{document}


\title{Quantum suppression law in a 3-D photonic chip\\ implementing fast Fourier transform\\ - Supplementary Information -}

\author{Andrea Crespi}
\affiliation{Istituto di Fotonica e Nanotecnologie, Consiglio Nazionale delle Ricerche (IFN-CNR), Piazza Leonardo da Vinci, 32,
I-20133 Milano, Italy}
\affiliation{Dipartimento di Fisica, Politecnico di Milano, Piazza Leonardo da Vinci, 32, I-20133 Milano, Italy}

\author{Roberto Osellame}
\email{roberto.osellame@polimi.it}
\affiliation{Istituto di Fotonica e Nanotecnologie, Consiglio
Nazionale delle Ricerche (IFN-CNR), Piazza Leonardo da Vinci, 32,
I-20133 Milano, Italy}
\affiliation{Dipartimento di Fisica, Politecnico di Milano, Piazza
Leonardo da Vinci, 32, I-20133 Milano, Italy}

\author{Roberta Ramponi}
\affiliation{Istituto di Fotonica e Nanotecnologie, Consiglio
Nazionale delle Ricerche (IFN-CNR), Piazza Leonardo da Vinci, 32,
I-20133 Milano, Italy}
\affiliation{Dipartimento di Fisica, Politecnico di Milano, Piazza
Leonardo da Vinci, 32, I-20133 Milano, Italy}

\author{Marco Bentivegna}
\affiliation{Dipartimento di Fisica, Sapienza Universit\`{a} di Roma,
Piazzale Aldo Moro 5, I-00185 Roma, Italy}

\author{Fulvio Flamini}
\affiliation{Dipartimento di Fisica, Sapienza Universit\`{a} di Roma,
Piazzale Aldo Moro 5, I-00185 Roma, Italy}

\author{Nicol\`o Spagnolo}
\affiliation{Dipartimento di Fisica, Sapienza Universit\`{a} di Roma,
Piazzale Aldo Moro 5, I-00185 Roma, Italy}

\author{Niko Viggianiello}
\affiliation{Dipartimento di Fisica, Sapienza Universit\`{a} di Roma,
Piazzale Aldo Moro 5, I-00185 Roma, Italy}

\author{Luca Innocenti}
\affiliation{Dipartimento di Fisica,  Universit\`{a} di Roma Tor Vergata,
Via della ricerca scientifica 1, I-00133 Roma, Italy}

\author{Paolo Mataloni}
\affiliation{Dipartimento di Fisica, Sapienza Universit\`{a} di Roma,
Piazzale Aldo Moro 5, I-00185 Roma, Italy}

\author{Fabio Sciarrino}
\email{fabio.sciarrino@uniroma1.it}
\affiliation{Dipartimento di Fisica, Sapienza Universit\`{a} di Roma,
Piazzale Aldo Moro 5, I-00185 Roma, Italy}

\maketitle

\clearpage

\section*{Interferometer design in the $2^p$-modes case}

In the following, we will schematically discuss the procedure to design an interferometer by implementing the Fast Fourier Transform (FFT) in the general case of $m=2^p$ modes, for some integer $p$.
\begin{itemize}
\item Each optical mode (waveguide) $k \in [1,m]$ can be numbered according to its binary representation and can be associated to the set
\begin{equation*}
(b_1, b_2, ..., b_p )_k
\end{equation*}
where $b_i = -1$ or $+1$ if the $i$-th bit of the binary representation of k is 0 or 1 respectively. The ordering of the $b_i$s is intended to be from the most significant bit $b_1$ to the least significant bit $b_p$.
\item The $p$-element vector $(b_1, b_2, ..., b_p )$ represents the coordinates of the vertices of a $p$-dimensional hypercube in $\mathbb{R}^p$.
\item The optical FFT algorithm \cite{Barak2007} consists of $p$ steps. The $j$-th step connects, making them interact, all the couples of modes that differ only for the $j$-th bit. (Proper phase terms are also added and a final relabelling of the outputs is needed, but this is not relevant here).
\item The position of each waveguide in the cross-section plane can be defined by suitably projecting in two dimensions the vertices of the hypercube. Each step of the FFT corresponds to connecting, by directional couplers, the couples of modes that have to interact in that step.
In particular, such connections corresponds to edges of the hypercube with a given direction, all parallel to each other (note that the edges of a $p$-dimensional hypercube are placed along $p$ possible directions).
\item It is worth of noting that with this layout, in each step, modes that have to be connected by couplers are all at the same relative distance and the projection of the couplers on the plane are all parallel lines. This means that such couplers can be all identical (except for deformations that may be needed for introducing phase terms) and waveguides never cross. This avoids parasitic coupling between modes that should stay separated and unwanted differences in the optical paths of the mode connections.
\end{itemize}

\section*{Two-photon source}

The generation of two-photon states is performed by pumping a 2-mm long BBO crystal with a 392.5 nm wavelength pulsed laser, with average power of 750 mW. After generation in the crystal, the two photons at 785 nm are spectrally filtered by means of 3 nm interferential filters, and coupled into single mode fibers. Polarization compensator and spatial delay lines are used before injecting the photons into the integrated interferometers through a single mode fiber array. Photons collection and detection is performed with a multimode fiber array and avalanche photodiodes. Finally, an electronic acquisition system allows to record two photon coincidences between arbitrary pairs of detectors. Typical coincidence rates for each collision-free output combination with distinguishable photons amounted to $\sim 70-80$ Hz (for the 4-mode chip) and $\sim 10-20$ Hz (for the 8-mode chip). 

\begin{figure}[p]

\begin{tabular}{|m{3cm}|m{10cm}|}
\hline 
\textbf{Dimension} & \textbf{Planar projection of the hypercube}
\\ \hline 
$p = 1$ &
\begin{center}
\begin{tikzpicture}[scale=1]
\footnotesize \sf
\Vertex{1}
\Vertex[x=2,y=0]{2}
\Edge(1)(2)
\end{tikzpicture}
\end{center}
\\ \hline
$p = 2$ &
\begin{center}
\begin{tikzpicture}[scale=1]
\footnotesize \sf
\Vertex{2}
\Vertex[x=0,y=2]{1}
\Vertex[x=2,y=0]{4}
\Vertex[x=2,y=2]{3}
\Edges(1,2,4,3,1)
\end{tikzpicture}
\end{center}
\\ \hline
$p = 3$ &
\begin{center}
\begin{tikzpicture}[scale=1]
\footnotesize \sf
\Vertex{4}
\Vertex[x=-1,y=1]{3}
\Vertex[x=0,y=2]{2}
\Vertex[x=-1,y=3]{1}
\Vertex[x=2,y=0]{8}
\Vertex[x=1,y=1]{7}
\Vertex[x=2,y=2]{6}
\Vertex[x=1,y=3]{5}
\Edges(1,2,4,3,1)
\Edges(5,6,8,7,5)
\Edge(1)(5)
\Edge(2)(6)
\Edge(3)(7)
\Edge(4)(8)
\end{tikzpicture}
\end{center}
\\ \hline
$p = 4$ &
\begin{center}
\begin{tikzpicture}[scale=1]
\footnotesize \sf
\Vertex{8}
\Vertex[x=2,y=1]{7}
\Vertex[x=-1,y=2]{6}
\Vertex[x=1,y=3]{5}
\Vertex[x=0,y=4]{4}
\Vertex[x=2,y=5]{3}
\Vertex[x=-1,y=6]{2}
\Vertex[x=1,y=7]{1}
\Vertex[x=4,y=0]{16}
\Vertex[x=6,y=1]{15}
\Vertex[x=3,y=2]{14}
\Vertex[x=5,y=3]{13}
\Vertex[x=4,y=4]{12}
\Vertex[x=6,y=5]{11}
\Vertex[x=3,y=6]{10}
\Vertex[x=5,y=7]{9}
\Edges(1,2,4,3,1)
\Edges(5,6,8,7,5)
\Edges(9,10,12,11,9)
\Edges(13,14,16,15,13)
\Edges(1,9,13,5,1)
\Edges(2,10,14,6,2)
\Edges(3,11,15,7,3)
\Edges(4,12,16,8,4)
\end{tikzpicture}
\end{center}
\\
\hline

\end{tabular}

\caption{Planar projections of hypercubes of dimension $p$ up to 4. The numbered vertices correspond to the possible position of the waveguides in the cross-section of a 3D interferometer which implements the Fast Fourier Transform. Each connection between the vertices corresponds to a directional coupler in a given section of the interferometer.}

\end{figure}